\documentclass[lettersize,journal]{IEEEtran}
\usepackage{amsmath,amsfonts}
\usepackage{algorithmic}
\usepackage{algorithm}
\usepackage{array}
\usepackage[caption=false,font=normalsize,labelfont=sf,textfont=sf]{subfig}
\usepackage{textcomp}
\usepackage{stfloats}
\usepackage{url}
\usepackage{verbatim}
\usepackage{graphicx}
\usepackage{cite}
\usepackage{amssymb}
\usepackage{xcolor}
\usepackage{epsfig}
\usepackage{epstopdf} 
\usepackage{multirow}
\usepackage{arydshln}
\usepackage{booktabs}
\usepackage{makecell}
\begin{document}

\title{Multi-Tier UAV Edge Computing Towards Long-Term Energy Stability for Low Altitude Networks}

\author{Yufei Ye,~\IEEEmembership{Graduate Student Member,~IEEE}, Shijian Gao,~\IEEEmembership{Member,~IEEE}, Xinhu Zheng,~\IEEEmembership{Member,~IEEE}, and Liuqing Yang,~\IEEEmembership{Fellow,~IEEE}
\thanks{Part of this work has been presented at the International Conference on Wireless Communications and Signal Processing (WCSP), Chongqing, China, October 23-25, 2025 \cite{yye2025multitier}. }
\thanks{Yufei Ye is with the Intelligent Transportation Thrust, The Hong Kong University of Science and Technology (Guangzhou), Guangzhou 511458, China (e-mail: yye760@connect.hkust-gz.edu.cn).}
\thanks{Shijian Gao is with the Internet of Things Thrust, The Hong Kong University of Science and Technology (Guangzhou), Guangzhou 511458, China (e-mail: shijiangao@hkust-gz.edu.cn).}
\thanks{Xinhu Zheng and Liuqing Yang are with the Intelligent Transportation Thrust and Internet of Things Thrust, The Hong Kong University of Science and Technology (Guangzhou), Guangzhou 511458, China (e-mail: xinhuzheng@hkust-gz.edu.cn; lqyang@ust.hk).}
}



\maketitle

\begin{abstract}
The agile mobility of Unmanned Aerial Vehicles (UAVs) makes them ideal for low-altitude edge computing. This paper proposes a novel multi-tier UAV edge computing system where lightweight Low-Tier UAVs (L-UAVs) function as edge servers for vehicle users, supported by a powerful High-Tier UAV (H-UAV) acting as a backup server. The objective is to minimize task execution delays while ensuring the long-term energy stability of the L-UAVs, despite unknown future system states. To this end, the problem is decoupled using Lyapunov optimization, which adaptively balances the priorities of task delays and L-UAV energy cost based on their real-time energy states. An efficient vehicle to L-UAV matching scheme is designed, and the joint optimization problem for task assignment, computing resource allocation, and trajectory control of L-UAVs and H-UAV is then solved via a Block Coordinate Descent (BCD) algorithm. Simulation results demonstrate a reduction in L-UAV transmission energy of over 26\% and superior L-UAV energy stability compared to existing benchmarks.
\end{abstract}

\begin{IEEEkeywords}
Edge computing, low-altitude networks, unmanned aerial vehicles, energy stability.
\end{IEEEkeywords}

\section{Introduction}
\IEEEPARstart{W}{ith} the substantial progress of vehicular communication technologies, autonomous driving, and artificial intelligence in recent years, the ever-increasing intelligence level in the Internet of Vehicles (IoV) has given rise to a growing demand for computing power \cite{xcheng2022mmwave}. Mobile Edge Computing (MEC) has emerged as a promising paradigm to meet this demand, greatly improving support for resource-intensive onboard IoV applications \cite{ymao2017asurvey, wshu2024anadaptive}.

However, the terrestrial servers with fixed positions face limitations in service coverage and are prone to signal blockages, which hinders their adaptability to the spatiotemporal dynamics of computing demands in IoV \cite{pchen2026qos}. The rapid development of the low-altitude economy (LAE) has positioned unmanned aerial vehicles (UAVs) as key enablers \cite{yjiang2025integrated}, as their elevated positions and agile mobility provide better Line-of-Sight (LoS) channels, broader coverage, and enhanced spatiotemporal scheduling capabilities \cite{njoshi2025energy}. This potential has catalyzed a surge of research into UAVs, spanning domains such as Integrated Sensing and Communication (ISAC) \cite{kmeng2023throughput, xcheng2022integrated}, millimeter-wave communications \cite{yliu2024latency}, and logistics \cite{pdu2024ai}.

Motivated by these advances, recent works have studied on UAV-assisted edge computing for air-ground networks. A UAV-assisted highway platoon system was developed in \cite{pzhao2025task} to conserve system energy. Authors in \cite{xwang2025generative} deployed UAVs to assist base station in processing overloaded tasks with subtask dependencies and minimized UAV energy cost and task processing delay. The age of information (AoI) of tasks and server rental costs were minimized in \cite{jyan2024deep}. By optimizing UAV trajectory and task offloading decisions, the task delay was minimized in \cite{myan2024edge}. Task offloading and content caching decisions were jointly optimized in \cite{jbai2025the} to minimize task latency in a vehicle–UAV–MEC–cloud cooperative system. The authors in \cite{jwang2024anadaptive} ensured offloading reliability under the worst-case channel conditions. The channel allocation and task processing fineness were optimized in \cite{cyang2022learning} to minimize the total cost. In \cite{yliu2024mobile}, the task migration costs among UAVs were stabilized via Lyapunov optimization. The authors in \cite{zliao2024anadaptive} achieved energy consumption balance among UAVs while maximizing successful task completions. Although UAVs exhibit deployment flexibility and dynamic adaptability, their compact designs limit their battery capacity and computing power. The highly volatile computational workloads in IoV expose UAVs to the risk of rapid energy depletion and task deadline violation.

Some studies proposed high-altitude platforms (HAPs) mounted with more powerful servers and stabler energy supply to compensate for the constraints of lightweight UAVs. The authors in \cite{sli2024joint} jointly optimized multi-dimensional resource of UAVs and HAP to minimize total task delay. The total energy consumption of an HAP-UAV network was minimized in \cite{zjia2025distributionally} employing distributionally robust optimization to handle estimation errors of channel state information. Joint user association and resource allocation in hierarchical aerial computing were optimized in \cite{anabi2025joint} by a two-tier matching game and enhanced soft actor-critic algorithm to minimize system task delay and energy cost. In \cite{ywang2024computation}, authors minimized long-term task cost while meeting the queuing mechanism. A stochastic and convex optimization method was developed in \cite{ychen2024energy} to conserve user energy, while system energy cost was reduced in \cite{hli2024dynamic} through an online dynamic optimization method. An enhanced actor-critic framework was proposed in \cite{mmorshed2024joint} to accommodate joint optimization of discrete and continuous variables. The authors in \cite{zjia2023hierarchical} address UAV selection and HAP offloading decision problems to maximize amount of successfully processed data. In \cite{nwaqar2022computation}, a double deep Q-learning model was developed to minimize task delay and energy cost in HAP-assisted IoV and prevent overestimation of action values. Despite the enhanced computing capabilities and broader coverage of HAPs, their operational altitudes result in high transmission energy and delay for UAVs when sending data to HAPs, which further strains UAV energy reserves.

A few works examined tiered UAV edge computing in air-ground networks, typically involving one single UAV physically connected to a terrestrial cloud server \cite{hyuan2024cost, bliu2023computation} or multiple upper-tier UAVs \cite{xren2024joint} as compensatory servers. On the one hand, most studies have focused on optimizing the trajectories of single-tier UAVs. The joint optimization of UAV trajectories across multiple tiers to improve the connectivity among inter-tier UAVs, thereby reducing the communication overhead, has not yet been fully considered. On the other hand, they either employ fixed weights for the trade-off between task execution delay and UAV energy cost or preset an upper bound on UAV energy consumption. More adaptive energy utilization strategies for multi-UAV systems that are better suited to highly dynamic and uncertain IoV environments have not been sufficiently investigated.

In this paper, we propose an innovative low-altitude multi-tier UAV edge computing system termed LATUS, consisting of the vehicle users (task sources), the lightweight Low-Tier UAVs (L-UAVs), and a High-Tier UAV (H-UAV). Positioned closer to vehicle users, L-UAVs function as lightweight edge servers, handling a portion of the computational tasks. Given their limited computing resources and energy reserves, an H-UAV equipped with a more powerful server and a larger-capacity battery operates at a slightly higher altitude as a mobile compensatory server to support the L-UAVs by managing additional workloads. We aim to minimize the overall task execution delay while ensuring the long-term energy stability of L-UAVs. The main challenges lie in the unavailability of future system states at each time slot and the high dimensionality arising from multifaceted coupled optimization variables. To tackle these challenges, the Lyapunov optimization is first employed to decouple the original problem into a series of deterministic problems that can be solved online within each time slot, through which the adaptive balance between task execution delays and L-UAV energy consumption based on their real-time energy states is achieved, thereby ensuring their dynamic energy stability. To deal with the problem in each time slot, an efficient vehicle to L-UAV (V2LU) matching scheme is first designed, then the task assignments, computing resource allocation, and flight trajectories for both L-UAVs and H-UAV are jointly optimized by the proposed block coordinate descent (BCD) based algorithm. LATUS not only shortens the communication distance between the H-UAV and L-UAVs but also jointly optimizes their trajectories to enhance the connectivities of the H-UAV with multiple L-UAVs, both contributing to reduced L-UAV transmission energy and delays. The hierarchical system architecture also facilitates workload balancing among UAVs, thereby enhancing the ecological stability of the entire system. Simulation results corroborate that LATUS outperforms benchmarks in maintaining the energy stability and reducing transmission energy of L-UAVs while attaining comparable task delays. The impacts of various system parameters on its performance are also systematically examined. These highlight the system's effectiveness in enhancing energy efficiency and responsiveness in dynamic air-ground network environments. The main contributions of this work are summarized as follows:

\begin{enumerate}
    \item We propose a novel multi-tier UAV edge computing system that leverages a High-Tier UAV (H-UAV) to support lightweight Low-Tier UAVs (L-UAVs), minimizing task delay while ensuring long-term L-UAV energy stability.
    
    \item We develop an online Lyapunov optimization framework that dynamically balances the delay-energy trade-off based on the real-time energy state of each L-UAV, converting the long-term stochastic problem into tractable per-slot problems.
    
    \item We design an efficient joint resource allocation algorithm, which integrates V2LU matching with a block coordinate descent method. Extensive simulations demonstrate LATUS reduces L-UAV transmission energy by over 26\% and achieves superior delay-energy stability.
\end{enumerate}

The rest of this article is organized as follows. In Section II, we present the model of the proposed low-altitude multi-tier UAV edge computing system. Then we detail the formulation of original problem and Lyapunov optimization-based problem transformation in Section III. The problem solving algorithm and performance analysis are elaborated in Section IV. We illustrate and discuss the simulation results in Section V and finally conclude this paper in Section VI.


\section{System Model}

In this section, we present the proposed multi-tier UAV edge computing system model, which encompasses the network model, UAV movement model, communication model, computation model, and L-UAV energy harvesting model.

\subsection{Network Model}
\begin{figure}[h]
\centering
\includegraphics[width=3.0 in]{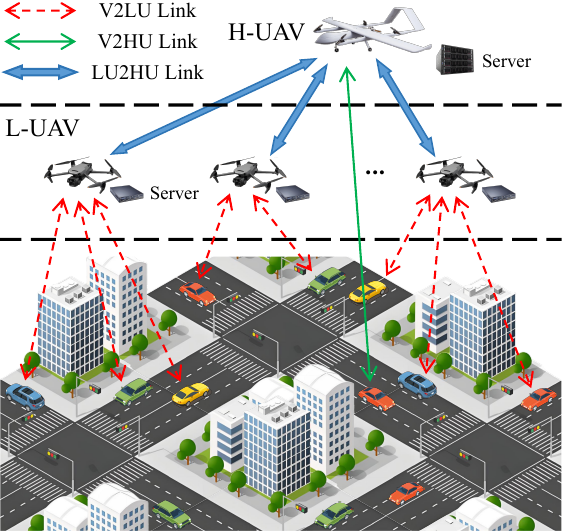}
\caption{Illustration of multi-tier UAV edge computing architecture in air-ground networks.}
\label{fig_1}
\end{figure}

Fig.~\ref{fig_1} depicts the proposed multi-tier UAV edge computing system, which comprises vehicle users in an urban traffic scenario, $U$ L-UAVs and one H-UAV. The L-UAVs serve as aerial lightweight servers that are closer to the ground vehicle users, while the H-UAV operates at a slightly higher altitude as an aerial backup server and central controller. The small-scale rotary-wing UAVs (RW-UAV) can be employed as L-UAVs, and the fixed-wing UAV (FW-UAV) \cite{xwang2023afixed} or electric Vertical Take-Off and Landing (eVTOL) aircraft \cite{hwei2024autonomous} with larger size can serve as the H-UAV, hence it can be equipped with a more powerful server and a battery with higher capacity. To capture the dynamics of the system, we divide each service period into $N$ time slots signified as $\mathcal{N}=\{0,1, ..., n, ...,N-1\}$. The length of each time slot is $\tau$. We denote the set of vehicles that request task offloading services in time slot $n$ as $\mathcal{V}(n)=\{1,2, ..., v, ...,V(n)\}$ and the set of L-UAVs as $\mathcal{U}=\{1,2, ..., u, ...,U\}$.

Without loss of generality, we denote the size of task data generated by vehicle $v$ in time slot $n$ as $D_v(n)$, with computational density $C_v(n)$ cycles/bit (i.e., the number of CPU cycles required to process each bit) and execution delay requirement $\tau_{v,max}(n)$. Each vehicle first transmits its initial task data to the associated L-UAV, which will compute a proportion $\alpha_v(n)$ of the task. The remaining part is then transmitted to the H-UAV for further computation. The horizontal positions of vehicle $v$ at time slot $n$ is signified by $\boldsymbol{G_v}(n) = (x_v(n), y_v(n))$ with vertical coordinate assumed to be 0 \cite{jwang2024anadaptive}. Given the small length of each time slot, the system within a single slot can be regarded as quasi-static. Hence, similar to \cite{sli2024joint, nwaqar2022computation}, the positions of vehicles and UAVs can be postulated to be constant within the same time slot.

\begin{table}[!t]
\caption{Main Notations and Definitions}
\label{tab:table1}
\centering
\setlength{\arrayrulewidth}{1pt} 
\begin{tabular}{!{\vrule width \arrayrulewidth} m{1.5cm}<{\centering} !{\vrule width \arrayrulewidth} m{6.2cm}<{\centering} !{\vrule width \arrayrulewidth}}
\Xhline{1pt}
\textbf{Notation} & \textbf{Definition}\\
\Xhline{1pt}
$\mathcal{U}, \mathcal{V}(n), \mathcal{N}$ & Set of L-UAVs, requesting vehicles at time slot $n$, and time slots, respectively\\
\Xhline{0.5pt}
$D_v, C_v$ & Size and computational density of the task data generated by vehicle $v$, respectively \\
\Xhline{0.5pt}
$\tau_{v,max}$ & Delay requirement of task generated by vehicle $v$\\
\Xhline{0.5pt}
$\alpha_v$ & Ratio of vehicle $v$'s task computed on its L-UAV\\
\Xhline{1pt}
$\boldsymbol{G_v},\boldsymbol{G_u},\boldsymbol{G_H}$ & Positions of vehicle $v$, L-UAV $u$, and H-UAV\\
\Xhline{0.5pt}
$H_1, H_2$ & Operating height of L-UAVs and H-UAV\\
\Xhline{0.5pt}
$d_{safe}$ & Minimum safety distance among L-UAVs\\
\Xhline{0.5pt}
$S_{u,max}$ & Maximum velocity of L-UAV $u$\\
\Xhline{0.5pt}
$S_{H,max}$ & Maximum velocity of H-UAV \\
\Xhline{1pt}
$R_{v,u}$ & Data rate between vehicle $v$ and L-UAV $u$ \\
\Xhline{0.5pt}
$R_{u,H}$ & Data rate between L-UAV $u$ and H-UAV \\
\Xhline{0.5pt}
$R_{v,H}$ & Data rate between vehicle $v$ and H-UAV\\
\Xhline{1pt}
$f_{v,u}$ & Computing resource allocated by L-UAV $u$ to the task of vehicle $v$ \\
\Xhline{0.5pt}
$f_{v,H}$ & Computing resource allocated by H-UAV to the task of vehicle $v$\\
\Xhline{1pt}
$T_{v,u}^{tr}$ & Data transmission delay from vehicle $v$ to L-UAV $u$ \\
\Xhline{0.5pt}
$T_{v,u,H}^{tr}$ & Data transmission delay from L-UAV $u$ to H-UAV\\
\Xhline{0.5pt}
$T_{v,H}^{tr}$ & Data transmission delay from vehicle $v$ to H-UAV \\
\Xhline{0.5pt}
$T_{v,u}^{comp}$ & Computation delay of vehicle $v$'s task on L-UAV $u$ \\
\Xhline{0.5pt}
$T_{v,H}^{comp}$ & Computation delay of vehicle $v$'s task on H-UAV\\
\Xhline{0.5pt}
$T_v$ & Total execution delay of vehicle $v$'s task\\
\Xhline{1pt}
$E_{v,u,H}^{tr}$ & Energy consumption of L-UAV $u$ for transmitting vehicle $v$'s task data to H-UAV\\
\Xhline{0.5pt}
$E_{v,u}^{comp}$ & Computation energy of L-UAV $u$ for vehicle $v$'s task\\
\Xhline{0.5pt}
$E_{u,flight}$ & Flight energy of L-UAV $u$\\
\Xhline{0.5pt}
$e_u$ & Amount of harvested energy by L-UAV $u$ \\
\Xhline{0.5pt}
$E_u$ & Total energy consumption of L-UAV $u$ \\
\Xhline{1pt}
$m_{v,u}$ & Matching indicator between vehicle $v$ and L-UAV $u$ \\
\Xhline{1pt}
$\tau$ & Duration of each time slot \\
\Xhline{1pt}
\end{tabular}
\end{table}

For the conciseness of presentation, the primary notations and the corresponding definitions in this work are summarized in Table~\ref{tab:table1}, where the index of time slot is omitted.

\vspace{-0.1cm}
\subsection{UAV Motion Model}

The horizontal position of L-UAV $u$ at time slot $n$ is symbolized as $\boldsymbol{G_u}(n) = (x_u(n), y_u(n))$ with a determined operating height $H_1$ \cite{jwang2024anadaptive, myan2024edge}. Consequently, the distance traveled by L-UAV $u$ between adjacent time slots can be expressed as $\Delta d_{u}(n)=\|\boldsymbol{G_u}(n)-\boldsymbol{G_u}(n-1)\|$. We denote the maximum velocity of L-UAV $u$ by $S_{u,max}$, hence it is subject to the maximum flight velocity constraint as
\begin{equation*}
\frac{\Delta d_{u}(n)}{\tau}\leq S_{u,max},\forall u \in \mathcal{U}, n \in \mathcal{N}. \tag{1}
\end{equation*}
Since multiple L-UAVs operate at the same altitude, it is imperative to ensure that the distance among them should be larger than the minimum safety distance denoted
by $d_{safe}$. Therefore, the positions of any two L-UAV $u$ and $u'$ must satisfy the following constraint:
\begin{equation*}
\| \boldsymbol{G_u}(n)-\boldsymbol{G_{u'}}(n)\|\geq d_{safe}, \forall u,u' \in \mathcal{U}, u\neq u', \forall n \in \mathcal{N}. \tag{2}
\end{equation*}
Similarly, we represent the horizontal position of H-UAV at time slot $n$ by $\boldsymbol{G_H}(n) = (x_H(n), y_H(n))$ with fixed operational height $H_2$. It should meet the maximum flight velocity constraint given by
\begin{equation*}
\frac{\Delta d_H(n)}{\tau}\leq S_{H,max},\forall n \in \mathcal{N}, \tag{3}
\end{equation*}
where $\Delta d_H(n)=\|\boldsymbol{G_H}(n)-\boldsymbol{G_H}(n-1)\|$ denotes the flight distance of H-UAV between two consecutive time slots.

We also consider the flight energy consumption of L-UAVs, which is related to their velocities. The average flight speed of L-UAV $u$ between time slot $n-1$ and $n$ can be expressed as $\overline{s}_u(n) = \frac{\Delta d_{u}(n)}{\tau}$. Hence the corresponding flight energy consumption of L-UAV $u$ is given by \cite{myan2024edge}
\begin{equation*}
E_{u,flight}(n) = 0.5M_u \tau {\overline{s}_u(n)}^2, \tag{4}
\end{equation*}
where $M_u$ is the mass of L-UAV $u$.

\vspace{-0.2cm}
\subsection{Communication Model}

The agile mobility of UAVs flying in the air enables them to effectively avoid obstacles and establish communication links with terrestrial vehicles. Hence, similar to \cite{pzhao2025task} and \cite{myan2024edge}, the wireless links among vehicles, L-UAVs, and H-UAV are considered to be dominated by Line-of-Sight (LoS) channels. The channel gain between vehicle $v$ and L-UAV $u$ is
\begin{equation*}
h_{v,u}(n)=\frac{\gamma_0}{d_{v,u}(n)^2}=\frac{\gamma_0}{H_1^2+\| \boldsymbol{G_u}(n) - \boldsymbol{G_v}(n)\|^2}, \tag{5}
\end{equation*}
where $d_{v,u}(n)$ is the distance between them at time slot $n$, $\gamma_0$ denotes the reference channel gain at distance of 1m. We employ Orthogonal Frequency Division Multiple Access (OFDMA) technology \cite{zjia2023hierarchical, ywang2024computation}, hence the transmission data rate between vehicle $v$ and L-UAV $u$ is expressed as
\begin{equation*}
R_{v,u}(n) = B_{v,u}(n) \log_2 \left(1+\frac{P_v(n)h_{v,u}(n)}{N_0 B_{v,u}(n)}\right), \tag{6}
\end{equation*}
where $B_{v,u}(n)$ denotes the allocated bandwidth for this V2LU channel, $P_v(n)$ is the transmit power of vehicle $v$ at time slot $n$, and $N_0$ is noise power spectral density. The corresponding data transmission delay is
\begin{equation*}
T_{v,u}^{tr}(n) = \frac{D_v(n)}{R_{v,u}(n)}. \tag{7}
\end{equation*}

In terms of L-UAV to H-UAV (LU2HU) communications, the channel gain and the transmission data rate between L-UAV $u$ and H-UAV are as following, respectively:
\begin{equation*}
h_{u,H}(n) = \frac{\gamma_0}{(H_2-H_1)^2+\| \boldsymbol{G_H}(n) - \boldsymbol{G_u}(n)\|^2}, \notag
\end{equation*}
\begin{equation*}
R_{u,H}(n) = B_{u,H}(n) \log_2 \left(1+\frac{P_u(n)h_{u,H}(n)}{N_0 B_{u,H}(n)}\right), \tag{8}
\end{equation*}
where $B_{u,H}(n)$ and $P_u(n)$ are the assigned bandwidth for this link and transmit power of L-UAV $u$, respectively. Accordingly, we can obtain the delay incurred by L-UAV $u$ in transmitting the remaining task data of vehicle $v$ to H-UAV as
\begin{equation*}
T_{v,u,H}^{tr}(n) = \frac{D_v(n)(1-\alpha_v(n))}{R_{u,H}(n)}, \tag{9}
\end{equation*}
where $\alpha_v(n)$ refers to the ratio of vehicle $v$'s task that is computed on its L-UAV. The energy consumed by L-UAV $u$ for transmitting vehicle $v$'s task data  to H-UAV is given by
\begin{equation*}
E_{v,u,H}^{tr}(n) = P_u(n)\cdot T_{v,u,H}^{tr}(n). \tag{10}
\end{equation*}

To further save the relay energy consumption of L-UAVs, the vehicles whose task computation ratios on their affiliated L-UAVs are zero directly offload their tasks to H-UAV via the vehicle to H-UAV (V2HU) link. The transmission data rate and delay from such vehicle $v$ to H-UAV are $R_{v,H}(n) = B_{v,H}(n) \log_2 \left(1+\frac{P_v(n)h_{v,H}(n)}{N_0 B_{v,H}(n)}\right)$ and $T_{v,H}^{tr}(n) = \frac{D_v(n)}{R_{v,H}(n)}$, respectively, where $B_{v,H}(n)$ is the bandwidth for V2HU direct link and $h_{v,H}(n) = \frac{\gamma_0}{H_2^2+\| \boldsymbol{G_H}(n) - \boldsymbol{G_v}(n)\|^2}$ expresses the corresponding channel gain. In general, the size of result data is significantly smaller than that of initial task data. Thus, similar to \cite{zjia2023hierarchical, bliu2023computation}, the delay and energy for result data transmission are neglected.

\vspace{-0.2cm}
\subsection{Computation Model}

We signify the computing resource (i.e., CPU frequency) allocated by L-UAV $u$ to vehicle $v$'s task at time slot $n$ as $f_{v,u}(n)$. Thus, the computation delay of this part can be expressed as
\begin{equation*}
T_{v,u}^{comp}(n) = \frac{D_v(n)C_v(n)\alpha_v(n)}{f_{v,u}(n)}. \tag{11}
\end{equation*}
The corresponding computation energy consumed by L-UAV $u$ is given by
\begin{equation*}
E_{v,u}^{comp}(n) = \kappa_u D_v(n)C_v(n)\alpha_v(n)f_{v,u}(n)^2, \tag{12}
\end{equation*}
where $\kappa_u$ denotes the energy efficiency factor \cite{jwang2024anadaptive} depending on the chip structure of L-UAV $u$’s processor. The computation delay for the H-UAV to process the remaining portion of the vehicle $v$'s task is
\begin{equation*}
T_{v,H}^{comp}(n) = \frac{D_v(n)C_v(n)(1-\alpha_v(n))}{f_{v,H}(n)}, \tag{13}
\end{equation*}
where $f_{v,H}(n)$ denotes the computing resource allocated by H-UAV to this task.

Based on above establishment, the total execution delay for each vehicle's task is the sum of V2LU data transmission delay, L-UAV computation delay, LU2HU data transmission delay, and H-UAV computation delay, which is expressed as
\begin{equation*}
T_v(n) = T_{v,u}^{tr}(n) + T_{v,u}^{comp}(n) + T_{v,u,H}^{tr}(n) + T_{v,H}^{comp}(n). \tag{14}
\end{equation*}
Note that for the vehicles that directly offload their task data to H-UAV, the first three delay terms should be replaced by $T_{v,H}^{tr}(n)$. The total energy consumed by L-UAV $u$ for its associated vehicle $v$ in time slot $n$ is
\begin{equation*}
E_{v,u}(n) =  E_{v,u}^{comp}(n) + E_{v,u,H}^{tr}(n). \tag{15}
\end{equation*}
Furthermore, the total energy consumption of L-UAV $u$ in time slot $n$ is expressed as
\begin{equation*}
E_u(n) = \sum_{v \in \mathcal{V}_u(n)} E_{v,u}(n) + E_{u,flight}(n), \tag{16}
\end{equation*}
where $\mathcal{V}_u(n)$ represents the set of vehicles served by L-UAV $u$ in time slot $n$.


\section{Problem Formulation}

In this section, we first formulate the primary optimization problem. Subsequently, we elaborate on the transformation from the original problem to an array of online problems in each time slot utilizing Lyapunov optimization.

\subsection{Problem Description}

Our objective is to minimize the overall task execution delay $T(n) = \sum\nolimits_{v \in \mathcal{V}(n)} T_v(n)$ of the system under the long-term L-UAV energy stability constraint and system resource restrictions through jointly optimizing the vehicle to L-UAV matching, task assignment ratios, computing resource allocation and trajectories of L-UAVs and H-UAV. 

Due to the limited energy storage, it is challenging for small-scale L-UAVs to provide continuous services for computation-intensive tasks. To prolong their endurance, we apply energy harvesting (EH) technology to L-UAVs, which enables them to exploit renewable energy from natural environment, such as solar, wind, and thermal energy, etc \cite{gkpandey2025uav}. The EH process can be modeled as the acquisition of consecutively arriving energy packets in each time slot, and they can be utilized in the subsequent time slots. We denote the amount of harvested energy by L-UAV $u$ at the beginning of time slot $n$ as $e_u (n)$. Given that this is an intermittent and highly stochastic process, to reflect this characteristic, $e_u (n)$ can be modeled as independent and identically distributed (i.i.d.) random variables with a maximum harvestable energy of $e_m$ \cite{zyang2022dynamic}.

The total energy of each L-UAV when fully charged at the initial time slot is symbolized as $E_{full}$, thereby establishing its reference energy quota for each time slot as $E_q = \frac{E_{full}}{N}$. Consequently, each L-UAV $u$ should adhere to the following long-term energy stability constraint:
\begin{equation*}
\frac{1}{N} \sum_{n=0}^{N-1} \mathbb{E}[E_u(n)-e_u(n)] \leq E_q, \forall u \in \mathcal{U}. \tag{17}
\end{equation*}
Hence, the primary optimization problem can be mathematically formulated as

$\mathcal{P}_0$: \textbf{Original Problem}
\begin{align*}
\min _{\substack {\boldsymbol{M}, \boldsymbol{f_U}, \boldsymbol{f_H} \\ \boldsymbol{\alpha}, \boldsymbol{G_U}, \boldsymbol{G_H}}}
&\frac{1}{N}\sum_{n=0}^{N-1} \mathbb{E} [T(n)]\\
\text {s.t. } \hspace{0.2cm} &0\leq\alpha_{v}(n)\leq1,\forall v, n, \tag{18a}\\
& f_{v,u}(n)\geq0,f_{v,H}(n)\geq0,\forall v,u,n, \tag{18b}\\
& \sum \nolimits_{v \in \mathcal{V}_u(n)} f_{v,u}(n)\leq F_{u},\forall u,n, \tag{18c}\\
& \sum \nolimits_{v \in \mathcal{V}(n)}f_{v,H}(n)\leq F_H,\forall n,\tag{18d}\\
& T_{v}(n)\leq\tau_{v,max},\forall v,n, \tag{18e}\\
& \text{(1)-(3), and (17)}.
\end{align*}

As for the optimization variables, $\boldsymbol{M} = \{ m_{v,u}(n) | v\in\mathcal{V}(n), u \in\mathcal{U}, n\in\mathcal{N} \}$ contains vehicle to L-UAV matching strategies, where $m_{v,u}(n)=1$ if vehicle $v$ is matched with L-UAV $u$ (i.e., $v \in \mathcal{V}_u(n)$), otherwise, $m_{v,u}(n)=0$. The remaining variable sets are specified as follows: $\boldsymbol{f_U} = \{ f_{v,u}(n)| v\in\mathcal{V}_u(n), u \in\mathcal{U}, n\in\mathcal{N}\}$, $\boldsymbol{f_H} = \{ f_{v,H}(n)| v\in\mathcal{V}(n), n\in\mathcal{N}\}$, $\boldsymbol{\alpha} = \{ \alpha_v(n)| v\in\mathcal{V}(n), n\in\mathcal{N}\}$, $\boldsymbol{G_U} = \{ \boldsymbol{G_u}(n)| u \in\mathcal{U}, n\in\mathcal{N}\} $, and $\boldsymbol{G_H} = \{ \boldsymbol{G_H}(n)| n\in\mathcal{N}\} $.

In terms of constraints, (18a) limits the range of task assignment ratios, while (18b) guarantees the nonnegativity of the computing resources allocated by L-UAVs and H-UAV. (18c) and (18d) ensure the total computing resources allocated by each L-UAV and H-UAV do not exceed their available capacities, respectively. The satisfaction of the delay requirement for each task is guaranteed by (18e). (1)-(3) are the UAV movement constraints discussed in Section II-B. The long-term energy stability of L-UAVs is ensured by (17).

\emph{Remark:} A significant challenge to solve this problem is that the system information such as the number and positions of vehicles, task profiles, and harvestable energy of L-UAVs for all time slots are required. However, the future system states are unknown at a specific time slot.

\vspace{-0.2cm}
\subsection{Problem Transformation}

To maintain the long-term energy stability of L-UAVs, we leverage the Lyapunov optimization technique \cite{zliu2025dnn} to transform the original problem into a series of deterministic problems, which can be solved in an online manner in each time slot without the need for information about future system.

To minimize the overall task delay while ensuring energy stability of L-UAVs over long-term time span, we employ the drift-plus-penalty approach \cite{ychen2024energy}. As for the objective function, the time-averaged task delay is suitable to be decoupled as a penalty term into the optimization target of each time slot. In order to decouple the long-term energy stability constraint of L-UAVs into each time slot, we establish energy quota deviation queues $\mathcal{Q}(n) = \{ Q_1(n), \ldots, Q_u(n), \ldots, Q_U(n)\}$ for L-UAVs. Specifically, the energy deviation queue of L-UAV $u$ at the beginning of time slot $n+1$ is expressed as
\begin{equation*}
Q_u(n+1) = \max\{ Q_u(n)+E_u(n)-e_u(n)-E_q , 0 \}, \tag{19}
\end{equation*}
where $E_u(n)$ and $e_u(n)$ denote the energy consumption and the available harvested energy of L-UAV $u$ in the previous time slot $n$, respectively. Thus, they can indicate the cumulative deviation from the energy quota for each L-UAV at the beginning of each time slot. The deviation queue at the initial time slot is $Q_u(0)=0, \forall u \in \mathcal{U}$. Based on these, the quadratic Lyapunov function is defined as $L(\mathcal{Q}(n)) = \frac{1}{2} \sum_{u=1}^{U} Q_u(n)^2$,
whose function value is non-negative and a smaller value indicates a lower degree of energy deviation for L-UAVs. Then, we define the Lyapunov drift function as
\begin{equation*}
\Delta L(\mathcal{Q}(n)) = \mathbb{E}[L(\mathcal{Q}(n+1)) - L(\mathcal{Q}(n))| \mathcal{Q}(n)], \tag{20}
\end{equation*}
which reflects the change of energy deviation queues between adjacent time slots. It serves as a key indicator for the energy stability of L-UAVs. We minimize the upper bound of the drift-plus-penalty function, which is given by
\begin{equation*}
\Delta_K(\mathcal{Q}(n)) = \Delta L(\mathcal{Q}(n)) + K \cdot \mathbb{E}[T(n) | \mathcal{Q}(n)] , \tag{21}
\end{equation*}
where $K > 0$ is a control parameter balancing the Lyapunov drift and penalty. We derive the upper bound of this function in the following proposition.

\textit{Proposition 1: For any feasible solution and any queue state, the drift-plus-penalty function is upper bounded by}
\begin{align*}
& \Delta_K(\mathcal{Q}(n)) \leq B + K \cdot \mathbb{E}[T(n) | \mathcal{Q}(n)] \\ &+ \sum_{u=1}^U \mathbb{E} [Q_u(n)(E_u(n)-e_u(n)-E_q)|\mathcal{Q}(n) ], \tag{22}
\end{align*}
\textit{where constant $B = \frac{1}{2} \sum_{u=1}^{U} [(E_u^{max})^2 + (e_m + E_q)^2]$ and $E_u^{max}$ is the upper limits for L-UAV $u$'s energy consumption.}

\emph{Proof:} According to (19), we have $Q_u(n+1)^2 \leq (Q_u(n)+E_u(n)-e_u(n)-E_q)^2$. Summing over all $U$ queues and dividing both sides by 2 yields
\begin{align*}
& \frac{1}{2} \sum_{u=1}^U Q_u(n+1)^2 \leq  \frac{1}{2} \sum_{u=1}^U Q_u(n)^2 + \frac{1}{2} \sum_{u=1}^U(E_u(n) \\ & -e_u(n)-E_q)^2 + \sum_{u=1}^U Q_u(n)(E_u(n)-e_u(n)-E_q). \tag{23}
\end{align*}
Combining the above inequality with (20), we obtain
\begin{align*}
&\Delta L(\mathcal{Q}(n)) \leq \frac{1}{2} \sum_{u=1}^U \mathbb{E}[(E_u(n)-e_u(n)-E_q)^2| \mathcal{Q}(n)] \\
& + \sum_{u=1}^U \mathbb{E}[Q_u(n) (E_u(n)-e_u(n)-E_q)| \mathcal{Q}(n)] \\
&\leq B + \sum_{u=1}^U \mathbb{E}[Q_u(n) (E_u(n)-e_u(n)-E_q)| \mathcal{Q}(n)], \tag{24}
\end{align*}
where the definition of $B$ is as given above and it is straightforward to demonstrate that $B \geq \frac{1}{2} \sum_{u=1}^U \mathbb{E}[(E_u(n)-e_u(n)-E_q)^2| \mathcal{Q}(n)]$. By adding the penalty term to both sides of the above inequality, we obtain (22). This completes the proof.
$\hfill\blacksquare$

After removing constant terms in the upper bound of drift-plus-penalty function, the optimization problem within each time slot is transformed into

$\mathcal{P}_1$: \textbf{Transformed Problem}
\begin{align*}
\min _{\substack {\boldsymbol{M}, \boldsymbol{f_U}, \boldsymbol{f_H} \\ \boldsymbol{\alpha}, \boldsymbol{G_U}, \boldsymbol{G_H}}}
& K \cdot T(n)+\sum_{u=1}^UQ_u(n) \cdot E_u(n)\\
\text {s.t. } \hspace{0.2cm} & \text{(18a)-(18e), (1)-(3)}.
\end{align*}

It can be observed that the objective function of the transformed problem is a dynamic weighted sum of the total task delay and the energy consumption of L-UAVs, with the $Q_u(n)$ of each L-UAV $u$ serving as variable weight for its energy consumption. Consequently, the L-UAV with a larger $Q_u(n)$ implies a greater deviation of its energy consumption from quota at the beginning of the current time slot, thus increasing the effort to conserve its energy consumption. Conversely, with small or zero $Q_u(n)$, we put greater emphasis on reducing task delay. In this way, we achieve the goal of dynamically adjusting the tradeoff between task execution delay and energy consumed by L-UAVs based on their real-time energy status.


\section{Joint Task Assignment, Resource Allocation, and UAV Trajectory Optimization}

In this section, we elaborate on the problem solution and subsequently conduct the complexity and performance analysis. The transformed problem is a non-convex mixed-integer nonlinear programming (MINLP) problem, which is difficult to directly obtain a globally optimal solution within polynomial time. We first design an efficient vehicle to L-UAV matching algorithm to determine the task offloading decision for each vehicle. Then the remaining problem is solved through a block coordinate descent (BCD)-based method.

\subsection{Vehicle to L-UAV Matching Algorithm}

\begin{figure}[!t]
\centering
\includegraphics[width=3.4 in]{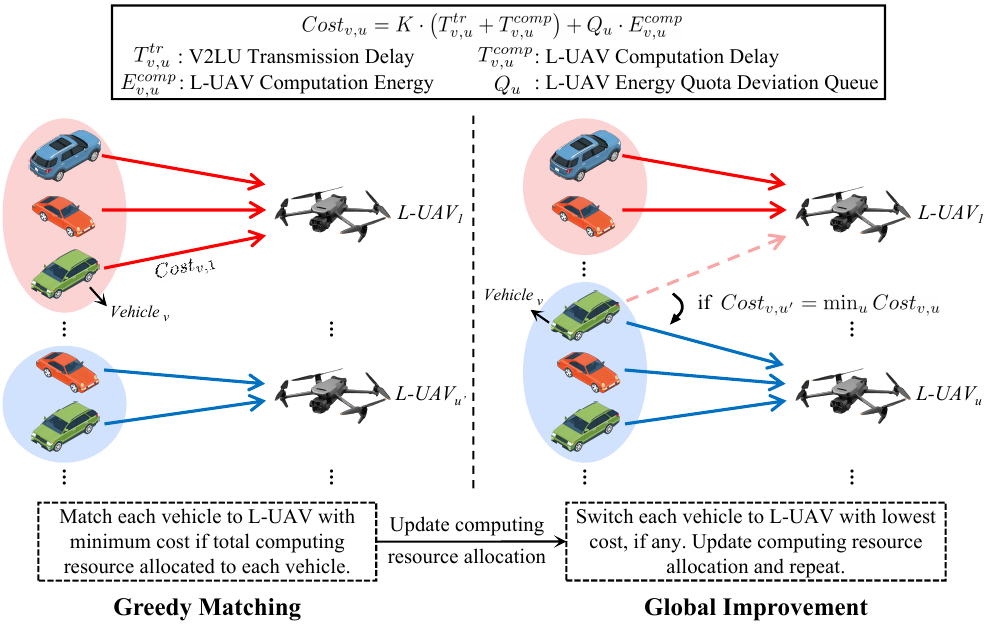}
\caption{The two-stage matching scheme between vehicles and L-UAVs.}
\label{matching}
\end{figure}

This matching process determines the task offloading decision for each vehicle, which serves as the prerequisite for subsequent optimizations. At this stage, neither the optimal computing resource allocation strategies for the two types of UAVs nor their trajectories are known. Therefore, we aim at obtaining a matching that can minimize the objective function to the greatest extent possible at the current stage. For the sake of brevity, the time slot index is omitted here without causing ambiguity. To achieve this goal, we first define the cost function for the task of each vehicle if it is offloaded to L-UAV $u$ as
\begin{equation}
Cost_{v,u} = K \cdot \left( T_{v,u}^{tr} + T_{v,u}^{comp}  \right) + Q_u \cdot E_{v,u}^{comp} , \tag{25}
\end{equation}
where we make a reasonable assumption that the H-UAV is temporarily excluded and all tasks are computed by L-UAVs. This process is illustrated as Fig.~\ref{matching}. The matching scheme consists of two stages: 
\begin{itemize}
    \item \emph{Stage 1 - Greedy matching:} We first initialize each matching indicator $m_{v,u}$ to zero and calculate the data rate for each pair of vehicle and L-UAV based on the input system parameters. In this stage, we assume that each L-UAV utilizes its entire available computing resource to compute each task. Subsequently, we calculate the initial cost for each vehicle's task if offloaded to each L-UAV based on (7), (11), (12), and (25). Afterwards, we greedily assign each task to the L-UAV associated with the minimum cost, thereby obtaining a temporary matching.
    \item \emph{Stage 2 - Global improvement:} According to the current matching result, each L-UAV updates its computing resource allocation strategy $f_u$ by evenly distributing resource among the vehicles it serves \cite{ywang2024computation}. It should be noted that for L-UAVs that are currently not assigned any vehicles, we set $f_u=F_u$. Accordingly, the cost of each task is recalculated. Based on the updated costs, we adjust the association for each vehicle by switching it to another L-UAV with a lower cost (if available). This process continues until no vehicle changes its association.
\end{itemize}

\textbf{Algorithm 1} summarizes the detailed steps. This method produces a high-quality, real-time matching that is dynamically adaptive and aligned with the overall system objective. Unlike pairwise vehicle-swapping methods, our approach adjusts the number of vehicles per L-UAV, leading to superior workload balancing and lower system cost.

\begin{algorithm}[!t]
    \caption{Vehicle to L-UAV Matching Algorithm}
    \label{alg:V2LUMatching}
    \renewcommand{\algorithmicrequire}{\textbf{Input:}}
    \renewcommand{\algorithmicensure}{\textbf{Output:}}
    
    \begin{algorithmic}[1]
        \REQUIRE $\mathcal{V}$, $\mathcal{U}$, $D_v$, $C_v$, $P_v$, $F_u$, $Q_u$, $\kappa_u$, $H_1$, $B_{v,u}$ at current time slot, and $\boldsymbol{G_v}$, $\boldsymbol{G_u}$ at previous time slot, $\forall v,u$   
        \ENSURE Vehicle to L-UAV matching result $\boldsymbol{M}$ at time slot $n$    
        
        \STATE  Initialize $\boldsymbol{M}$ with $m_{v,u}=0, \forall v,u$.

        \FOR{$v \in \mathcal{V}$}
            \FOR{$u \in \mathcal{U}$}
                \STATE Calculate $T_{v,u}^{tr}$, $T_{v,u}^{comp}$, and $E_{v,u}^{comp}$ based on (7), $T_{v,u}^{comp}=D_v C_v/F_u$, and $E_{v,u}^{comp}=\kappa_u D_v C_v F_u^2$, respectively.
                \STATE Calculate the $Cost_{v,u}$ for each vehicle $v$'s task if offloaded to L-UAV $u$ based on (25).
                
            \ENDFOR
        \ENDFOR

        \STATE Assign each vehicle to the L-UAV $u$ with the minimum cost temporarily (i.e., set corresponding $m_{v,u}=1$).
        
        \REPEAT
            \FOR{$u \in \mathcal{U}$}
                \STATE $f_u \leftarrow F_u / \max \{ \sum_{v \in \mathcal{V}} m_{v,u}, 1\}$
                \FOR{$v \in \mathcal{V}$}
                    \STATE Calculate their $Cost_{v,u}$ with updated $f_u$.
                \ENDFOR
            \ENDFOR
            \FOR{$v \in \mathcal{V}$}
                \STATE $Current\_cost_v \leftarrow \sum_{u \in \mathcal{U}} m_{v,u} Cost_{v,u}$

                \STATE Switch vehicle $v$'s association to L-UAV $u'$ with the lowest $Cost_{v,u'}$ if $Cost_{v,u'} \neq Current\_cost_v$.
                
            \ENDFOR
        \UNTIL No vehicle changes its association with L-UAV.
        
        \RETURN Matching result $\boldsymbol{M}$
    \end{algorithmic}
\end{algorithm}

\subsection{Joint Optimization of Task Assignment Ratio and UAV Computing Resource Allocation}

The remaining problem is still non-trivial, involving five types of optimization variables. We can elegantly decompose it into several low-dimensional and tractable subproblems, thereby reducing the solving complexity. This naturally motivates the adoption of BCD method, which is well-suited for large-scale optimization.

The core idea of the BCD method is to decompose the complex problem into multiple subproblems by reasonably grouping the optimization variables, which enables us to design suitable solution tailored to their distinct structures. Each subproblem is then iteratively optimized to ultimately converge into an efficient solution of original problem. Based on this fundamental idea, we decompose the remaining problem into three subproblems: joint optimization of task assignment ratios and computing resource allocation for L-UAVs and H-UAV, trajectory optimization of L-UAVs and that of H-UAV. 

The ratio of each vehicle's task assigned to the two types of UAVs and the corresponding computing resource allocated by them are strongly correlated and exhibit high sensitivity to the numerical changes of each other. Therefore, to reduce the number of overall iterations, thereby improving the convergence efficiency, we group these three categories of variables together for joint optimization. Given the V2LU matching and trajectories of the two types of UAVs and retaining only the related terms in objective function and constraints, this subproblem can be formulated as

$\mathcal{P}_2$: \textbf{Task Assignment and Computing Resource Allocation}
\begin{align*}
\min _{\substack {\boldsymbol{\alpha}, \boldsymbol{f_U}, \boldsymbol{f_H}}}
&K \cdot \sum_{v=1}^{V(n)} T_v^{comp}(n) + \sum_{u=1}^U Q_u(n) \cdot E_u(n)\\
\text {s.t. } \hspace{0.2cm} & \text{(18a)-(18e)},
\end{align*}
where $T_v^{comp}(n)=T_{v,u}^{comp}(n) + T_{v,u,H}^{tr}(n) + T_{v,H}^{comp}(n)$. For notational simplicity, the time slot index $n$ is omitted in the remainder of this subsection.

We denote the objective function of Subproblem-1 as $F$. When $\alpha_v$ for all vehicles are determined, the Hessian matrix of function $F$ w.r.t. $f_{v,u}$ and $f_{v,H}$ is expressed as
\begin{equation*}
\boldsymbol{H} = 
\begin{bmatrix}
H_{11}&H_{12}\\H_{21}&H_{22}
\end{bmatrix} = 
\begin{bmatrix}
\frac{\partial^2 F}{\partial f_{v,u}^2} & \frac{\partial^2 F}{\partial f_{v,u} \partial f_{v,H}}\\
\frac{\partial^2 F}{\partial f_{v,H} \partial f_{v,u}} & \frac{\partial^2 F}{\partial f_{v,H}^2}
\end{bmatrix} , \tag{26}
\end{equation*}
where \begin{small}$\frac{\partial^2 F}{\partial f_{v,u}^2} = \frac{2K \alpha_v D_v C_v}{f_{v,u}^3} + 2 \alpha_v \kappa_u Q_u D_v C_v \geq 0$\end{small}, \begin{small}$\frac{\partial^2 F}{\partial f_{v,H}^2} = \frac{2K (1-\alpha_v) D_v C_v}{f_{v,H}^3} \geq 0$
\end{small}, and \begin{small}$\frac{\partial^2 F}{\partial f_{v,u} \partial f_{v,H}} = \frac{\partial^2 F}{\partial f_{v,H} \partial f_{v,u}} = 0$
\end{small}. This indicates that the matrix $\boldsymbol{H}$ is positive semi-definite, thereby proving that the objective function is convex w.r.t. $f_{v,u}$ and $f_{v,H}$ when $\alpha_v$ is given. Similarly, constraint (18e) is convex with them. (18b)-(18d) are affine constraints of $f_{v,u}$ and $f_{v,H}$ that are also convex. Therefore, Subproblem-1 is a convex optimization problem with determined $\alpha_v$, which can be optimally solved by the standard convex optimization tools such as CVX solver \cite{cvx}. When $f_{v,u}$ and $f_{v,H}$ are given, the objective function is linear with $\alpha_v$, hence the optimal value can be found by the linear search-based method. Based on the above analysis, we utilize an alternating optimization approach to address these three types of variables. Specifically, we initially assign a feasible value to each $\alpha_v$, based on which we optimize $f_{v,u}$ and $f_{v,H}$. Subsequently, we update $\alpha_v$ using the once-optimized $f_{v,u}$ and $f_{v,H}$. We repeat those steps until convergence, thereby obtaining the final solution.

\subsection{Trajectory Optimization for L-UAVs and H-UAV}

In this subsection, with determined V2LU matching result, task assignment ratios, and computing resource allocation strategies of L-UAVs and H-UAV, we elaborate on the trajectory optimization for them.

As for L-UAVs, their trajectories are tightly coupled with communications to multiple served vehicles and their flight energy consumption. We optimize their trajectories to minimize the adaptive weighted sum of the overall V2LU data transmission delays and their flight energy according to the real-time energy status of each L-UAV. Consequently, the subproblem of trajectory optimization for L-UAVs is formulated as

$\mathcal{P}_3$: \textbf{L-UAV Trajectory Optimization}
\begin{align*}
\min _{\substack {\boldsymbol{G_U}(n)}}
& K \cdot \sum_{u=1}^U \sum_{v=1}^{V_u(n)} T_{v,u}^{tr}(n) + \sum_{u=1}^U Q_u(n) \cdot E_{u,flight}(n)\\
\text {s.t. } & \text{(1), (2), and (18e)}.
\end{align*}

The challenge lies in the fact that this is a non-convex problem, since the objective function, constraint (2) and (18e) are all non-convex with $\boldsymbol{G_u}(n)$. Further analysis reveals that the objective function to be minimized and the left-hand side of constraint (18e), which must be kept below a certain threshold, are well-suited for relaxation by convex upper bounds. Similarly, constraint (2) can also be satisfied by ensuring the convex lower bound of its left-hand side exceeds the minimum safety distance. Based on this insight, we adopt the successive convex approximation (SCA) technique to first transform this problem into a sequence of approximated convex problems and then continuously solve them until convergence.

In the objective function, $T_{v,u}^{tr}(n)$ is the non-convex term. Its denominator $R_{v,u}(n)$ is also a non-convex function of the optimization variable $\boldsymbol{G_u}(n)$, but it is convex w.r.t. the entire $\| \boldsymbol{G_u}(n) - \boldsymbol{G_v}(n)\|^2$. Define $\varphi_{v,u} (n) = \| \boldsymbol{G_u}(n) - \boldsymbol{G_v}(n)\|^2$, hence $R_{v,u}(n)$ can be globally lower-bounded by its first-order Taylor expansion with $\varphi_{v,u} (n)$ at any point. We denote $\boldsymbol{G_u^k}(n)$ as the position of L-UAV $u$ at the $k$-th iteration. Then the lower bound of $R_{v,u}(n)$ can be derived as
\begin{equation}
\widehat{R}_{v,u}(n)=R_{v,u}^k(n)+\nabla R_{v,u}^k(n)\left(\varphi_{v,u}(n)-\varphi_{v,u}^k(n)\right), \tag{27}
\end{equation}
where $R_{v,u}^k(n)$ and $\nabla R_{v,u}^k(n)$ signify the data transmission rate between vehicle $v$ and L-UAV $u$ and the first-order derivative of $R_{v,u}(n)$ w.r.t. $\varphi_{v,u} (n)$ at the $k$-th iteration, respectively. Specifically, they are given by
\begin{equation}
R_{v,u}^k(n) = B_{v,u}(n) \log_2 \left(1+\frac{P_v(n)\delta}{ H_1^2 + 
\varphi_{v,u}^k(n) }\right), \notag
\end{equation}
\begin{equation}
\nabla R_{v,u}^k(n) = \frac{- B_{v,u}(n)P_v(n) \delta \log_2e}{(H_1^2 + \varphi_{v,u}^k(n)) (H_1^2 + \varphi_{v,u}^k(n) + P_v(n) \delta)}, \tag{28}
\end{equation}
where $\varphi_{v,u}^k(n) = \| \boldsymbol{G_u^k}(n) - \boldsymbol{G_v}(n) \|^2$ and $\delta = \gamma_0 / N_0$. Consequently, we obtain a convex upper bound for $T_{v,u}^{tr}(n)$, which is given by $T_{v,u}^{tr}(n) \leq \frac{D_v(n)}{\widehat{R}_{v,u}(n)}$. Since $T_{v,u}^{tr}(n)$ is also contained in constraint (18e), following a similar derivation, a convex upper bound for the left-hand side of constraint (18e) can be expressed as $T_v(n) \leq \frac{D_v(n)}{\widehat{R}_{v,u}(n)} + T_{v}^{comp}(n)$.

As for the constraint (2), we rewrite it as $\| \boldsymbol{G_u}(n)-\boldsymbol{G_{u'}}(n)\|^2 \geq d_{safe}^2$. The left-hand side is non-convex with $\boldsymbol{G_u}(n)$ and $\boldsymbol{G_{u'}}(n)$, but convex with $\boldsymbol{G_u}(n)-\boldsymbol{G_{u'}}(n)$, which motivates us to gain its lower bound function by utilizing its first-order Taylor expansion at any point w.r.t. $\boldsymbol{G_u}(n)-\boldsymbol{G_{u'}}(n)$ calculated as
\begin{align*}
\widehat{d}_{u,u'}^2(n) = &  \|\boldsymbol{G}_{\boldsymbol{u}}^{\boldsymbol{k}}(n)-\boldsymbol{G}_{\boldsymbol{u}^{\prime}}^{\boldsymbol{k}}(n)\|^{2} +2\left(\boldsymbol{G}_{\boldsymbol{u}}^{\boldsymbol{k}}(n)-\boldsymbol{G}_{\boldsymbol{u}^{\prime}}^{\boldsymbol{k}}(n)\right)^{T} \\
&\left((\boldsymbol{G}_{\boldsymbol{u}}(n)-\boldsymbol{G}_{\boldsymbol{u}^{\prime}}(n))-(\boldsymbol{G}_{\boldsymbol{u}}^{\boldsymbol{k}}(n)-\boldsymbol{G}_{\boldsymbol{u}^{\prime}}^{\boldsymbol{k}}(n))\right). \tag{29}
\end{align*}

Through the above transformation, we convert the non-convex problem into its convex approximation problem as

$\mathcal{P}_3'$: \textbf{Convex Approximation Problem of $\mathcal{P}_3$}
\begin{align*}
\min _{\substack {\boldsymbol{G_U}(n)}}
& K \cdot \sum_{u=1}^U \sum_{v=1}^{V_u(n)} \frac{D_v(n)}{\widehat{R}_{v,u}(n)} + \sum_{u=1}^U Q_u(n) \cdot E_{u,flight}(n)\\
\text {s.t. } \hspace{0.1cm} & \text{(1),} \\
& \widehat{d}_{u,u'}^2(n) \geq d_{safe}^2, \forall u,u' \in \mathcal{U}, u\neq u', \tag{30a}\\
& \frac{D_v(n)}{\widehat{R}_{v,u}(n)} + T_{v}^{comp}(n) \leq \tau_{v,max}, \forall v, \tag{30b}
\end{align*}
which can be solved by CVX solver. The problem in each iteration is formed based on the solution from previous iteration. By iteratively solving them, it ultimately converges to an efficient near-optimal solution that satisfies all constraints.

Given the V2LU matching result, task assignment ratios, computing resource allocation strategy, and trajectories of L-UAVs, we further optimize H-UAV's trajectory to facilitate its collaboration with multiple L-UAVs based on their path design. This subproblem is expressed as

$\mathcal{P}_4$: \textbf{H-UAV Trajectory Optimization}
\begin{align*}
\min _{\substack {\boldsymbol{G_H}(n)}}
& \sum_{u=1}^U \left( K \cdot \sum_{v=1}^{V_u(n)} T_{v,u,H}^{tr}(n) + Q_u(n) \sum_{v=1}^{V_u(n)} E_{v,u,H}^{tr}(n) \right)\\
\text {s.t. } \hspace{0.1cm} & \text{(3) and (21e)},
\end{align*}
which indicates dual significances of H-UAV trajectory optimization: (1) reducing the data transmission delay from L-UAVs to H-UAV and (2) conserving the transmission energy for each L-UAV according to its real-time energy condition.

The terms $T_{v,u,H}^{tr}(n)$ and $E_{v,u,H}^{tr}(n)$ are non-convex w.r.t. $\boldsymbol{G_H}(n)$. Since the non-convexity of the latter stems exclusively from its dependence on the former, we only need to deal with $T_{v,u,H}^{tr}(n)$. Similar to the aforementioned analysis, we define $\varphi_{u,H}(n) = \| \boldsymbol{G_H}(n) - \boldsymbol{G_u}(n)\|^2$. Hence the convex upper bound of $T_{v,u,H}^{tr}(n)$ can be derived as $T_{v,u,H}^{tr}(n) \leq \frac{D_v(n) ( 1- \alpha_v(n) )}{\widehat{R}_{u,H}(n)}$, where $\widehat{R}_{u,H}(n)$ is given by
\begin{equation}
\widehat{R}_{u,H}(n) = R_{u,H}^i(n) + \nabla R_{u,H}^i(n)\left(\varphi_{u,H}(n)-\varphi_{u,H}^i(n)\right). \tag{31}
\end{equation}
In (31), $R_{u,H}^i(n)$ and $\nabla R_{u,H}^i(n)$ denote the data transmission rate from L-UAV $u$ to H-UAV and the first-order derivative of $R_{u,H}(n)$ w.r.t. $\varphi_{u,H}(n)$ at the $i$-th iteration, respectively. Consequently, the subproblem of H-UAV trajectory optimization is converted into the convex approximation problem as

$\mathcal{P}_4'$: \textbf{Convex Approximation Problem of $\mathcal{P}_4$}
\begin{align*}
\min _{\substack {\boldsymbol{G_H}(n)}}
& \sum_{u=1}^U \left( K + Q_u(n) P_u(n) \right) \sum_{v=1}^{V_u(n)} \frac{D_v(n) ( 1- \alpha_v(n) )}{\widehat{R}_{u,H}(n)}\\
\text {s.t. } \hspace{0.1cm} & \text{(3)}, \\
& T_{v,u}^{tr}(n) + T_{v,u}^{comp}(n) + \frac{D_v(n) ( 1- \alpha_v(n) )}{\widehat{R}_{u,H}(n)} \\
& + T_{v,H}^{comp} \leq \tau_{v,max}, \forall v, \tag{32}
\end{align*}
which can be iteratively solved by CVX solver. By this point, we complete trajectory optimization for L-UAVs and H-UAV.

\subsection{Complexity Analysis}

The steps of the overall algorithm are outlined in the \textbf{Algorithm 2}. To begin with, we initialize an array of system parameters and update energy quota deviation queue for each L-UAV. Then we first execute Algorithm 1 to obtain the vehicle to L-UAV matching result. Subsequently, since the subproblem $\mathcal{P}_2$ depends on the most recently optimized trajectories of the two types of UAVs and H-UAV trajectory optimization relies on the results of $\mathcal{P}_2$ and $\mathcal{P}_3'$, we iteratively solve $\mathcal{P}_2$, $\mathcal{P}_3'$, and $\mathcal{P}_4'$ in sequence, until convergence or the maximum number of iterations is reached. The superscript $j$ represents the current solution at the beginning of the $j$-th iteration.

\begin{algorithm}[!t]
    \caption{Joint Optimization Algorithm of Task Offloading, Computing Resource Allocation, and UAV Trajectories.}
    \label{alg:overall}
    \renewcommand{\algorithmicrequire}{\textbf{Input:}}
    \renewcommand{\algorithmicensure}{\textbf{Output:}}
    
    \begin{algorithmic}[1]
        \REQUIRE Vehicle set $\mathcal{V}(n)$, L-UAV set $\mathcal{U}$, task profile $D_v(n)$ and $C_v(n)$, vehicle positions $\boldsymbol{G_v}(n)$, computing capacities of L-UAVs $F_u$ and H-UAV $F_H$, positions of L-UAVs $\boldsymbol{G_u}(n-1)$ and H-UAV $\boldsymbol{G_H}(n-1)$.   
        \ENSURE Vehicle to L-UAV matching result $\boldsymbol{M}(n)$, task assignment ratios $\boldsymbol{\alpha}(n)$, L-UAV and H-UAV computing resource allocation $\boldsymbol{f_U}(n)$ and $\boldsymbol{f_H}(n)$, L-UAV and H-UAV trajectories $\boldsymbol{G_U}(n)$ and $\boldsymbol{G_H}(n)$.     
        
        \STATE Update $Q_u(n)$ according to (19) or set $Q_u(n) = 0$ at the initial time slot, and initialize iteration number $j=0$.
        \STATE Execute \textbf{Algorithm 1} and obtain $\boldsymbol{M}(n)$.

        \REPEAT
            \STATE Solve $\mathcal{P}_2$ to obtain $\boldsymbol{\alpha}^{j+1}$, $\boldsymbol{f_U}^{j+1}$, $\boldsymbol{f_H}^{j+1}$ with given $\boldsymbol{G_U}^j$ and $\boldsymbol{G_H}^j$.
            \STATE Solve $\mathcal{P}_3'$ to obtain $\boldsymbol{G_U}^{j+1}$ with given $\boldsymbol{\alpha}^{j+1}$, $\boldsymbol{f_U}^{j+1}$, $\boldsymbol{f_H}^{j+1}$, and $\boldsymbol{G_H}^j$.
            \STATE Solve $\mathcal{P}_4'$ to obtain $\boldsymbol{G_H}^{j+1}$ with given $\boldsymbol{\alpha}^{j+1}$, $\boldsymbol{f_U}^{j+1}$, $\boldsymbol{f_H}^{j+1}$, and $\boldsymbol{G_U}^{j+1}$.
            \STATE Update the value of objective function.
            \STATE Update $j=j+1$.

        \UNTIL The objective value converges or $j>j_{max}$.

    \end{algorithmic}
\end{algorithm}

In Algorithm 1, both the calculation of initial costs and the greedy assignment require traversing all vehicles and L-UAVs to calculate or identify the minimum cost. Thus, their complexities are both $\mathcal{O}\left(VU\right)$. Similarly, the global improvement stage undergoes such traversal twice in each round. Denote the number of improvement rounds as $I_1$, then the complexity of this stage is $\mathcal{O}\left( 2I_1VU \right)$. Hence the overall complexity of Algorithm 1 is $\mathcal{O}\left(VU + VU + 2I_1VU \right) \approx \mathcal{O} \left( I_1VU \right)$. Regarding the solution to problem $\mathcal{P}_2$, the complexity of interior-point method in CVX solver and the linear search method scale approximately as $\mathcal{O}((2V)^{3.5})$ and $\mathcal{O}(V^{3.5})$, respectively. Thus the complexity of solving problem $\mathcal{P}_2$ is $\mathcal{O}(I_2((2V)^{3.5}+V^{3.5})) \approx \mathcal{O}(I_2 V^{3.5})$, where $I_2$ denotes the iteration number. The complexity of solving problem $\mathcal{P}_3'$ and $\mathcal{P}_4'$ with SCA method are $\mathcal{O}((VU)^{3.5})$ and $\mathcal{O}(U^{3.5})$, respectively. Therefore, the overall complexity of the proposed method is $\mathcal{O}(J(I_1VU+I_2V^{3.5}+(VU)^{3.5}+U^{3.5})) \approx \mathcal{O}(J(VU)^{3.5})$, where $J$ signifies the iteration number.

\subsection{Performance Analysis}
In this subsection, we conduct theoretical performance analysis for the proposed method. We elaborate the performance gap between Lyapunov optimization-based solution and theoretically optimal solution of the original problem, and the energy stability of L-UAVs. This analysis will demonstrate that both the average system delay and the energy quota deviation queues are controllable within their respective upper bounds.

\textit{Proposition 2: For a given control parameter $K$, the time-average overall task delay satisfies}
\begin{equation}
\frac{1}{N}\sum_{n=0}^{N-1} \mathbb{E} [T(n)|\mathcal{Q}(n)] \leq T^* + \frac{B}{K}, \tag{33}
\end{equation}
\textit{where $T^*$ denotes the optimal overall task delay in the original problem and the definition of $B$ is consistent with (22). Formula (33) indicates that task delay of the proposed method monotonically approaches the optimal delay as $K$ increases. For sufficiently large $K$, the task delay converges arbitrarily close to the optimum.}

\emph{Proof:} We denote all the optimization variables collectively as $\boldsymbol{X}(n)$, and represent the overall task delay and L-UAV energy consumption under $\boldsymbol{X}(n)$ as $T(\boldsymbol{X}(n))$ and $E_u(\boldsymbol{X}(n))$, respectively. According to \cite{ztong2022dynamic}, we can rewrite (22) as
\begin{align*}
\Delta_K(\mathcal{Q}(n)) \leq & B + K \cdot \mathbb{E} [T(\boldsymbol{X}^*(n)) | \mathcal{Q}(n)] \\ +& \sum_{u=1}^U \mathbb{E} [Q_u(n)(E_u(\boldsymbol{X}^*(n))-e_u(n)-E_q)|\mathcal{Q}(n) ] \\
\leq & B + K \cdot T^*. \tag{34}
\end{align*}
Summing over $N$ time slots, then we obtain $L(\mathcal{Q}(N)) - L(\mathcal{Q}(0)) + K \sum_{n=0}^{N-1} \mathbb{E} [T(n)|\mathcal{Q}(n)] \leq BN + KNT^*$.
Since initial deviation queues are empty, we have $L(\mathcal{Q}(0))=0$. Given that $L(\mathcal{Q}(N))\geq0$ and dividing both sides by $KN$, we can derive (33). This completes the proof.
$\hfill\blacksquare$

\textit{Proposition 3: The upper bound of average energy quota deviation queues for L-UAVs is}
\begin{equation}
\frac{1}{N} \sum_{n=0}^{N-1} \sum_{u=1}^U \mathbb{E}[Q_u(n) | \mathcal{Q}(n)] \leq \frac{1}{\varepsilon}\left[ B + K(T^{max}-T^*) \right], \tag{35}
\end{equation}
\textit{where $T^{max}$ represents the maximum overall task delay.}

\emph{Proof:} A practically controllable system generally satisfies the following premise \cite{ztong2022dynamic}: there always exists a decision set $\boldsymbol{X}'(n)$ such that $\mathbb{E}[E_u(\boldsymbol{X}'(n))-e_u(n)-E_q] \leq - \varepsilon$ holds, where $\varepsilon$ is a positive real number slightly greater than zero. Combining the above with (22), we can obtain
\begin{align*}
\Delta_K(\mathcal{Q}(n)) \leq & B + K \cdot \mathbb{E}[T(\boldsymbol{X}'(n)) | \mathcal{Q}(n)]\\
& - \varepsilon \sum_{u=1}^U \mathbb{E} [Q_u(n)|\mathcal{Q}(n) ]. \tag{36}
\end{align*}
We replace $\mathbb{E}[T(n) | \mathcal{Q}(n)]$ on the left-hand side of inequality with $T^*$ and $\mathbb{E}[T(\boldsymbol{X}'(n)) | \mathcal{Q}(n)]$ on the right-hand side with $T^{max}$, then the inequality still holds. Summing over $N$ time slots and dividing by $\varepsilon N$ for both sides, we have
\begin{align*}
&\frac{1}{N} \sum_{n=0}^{N-1} \sum_{u=1}^U \mathbb{E} [Q_u(n)|\mathcal{Q}(n) ] \leq \\
&\frac{1}{\varepsilon}\left[ B + K(T^{max}-T^*) - \frac{\mathbb{E}[L(\mathcal{Q}(N)]}{N}\right]. \tag{37}
\end{align*}
Due to the nonnegativity of $L(\mathcal{Q}(N)$, after scaling the above inequality by removing term $\frac{\mathbb{E}[L(\mathcal{Q}(N)]}{N}$, we can derive (35). This completes the proof.
$\hfill\blacksquare$


\section{Simulation Results}

In this section, we conduct extensive simulation experiments to validate the superiority of the proposed method over other benchmarks and examine the impact of various system parameters on its performance.

We consider a 1 km $\times$ 1 km square traffic area, where vehicles are randomly distributed, with the speed ranging from 30 to 80 km/h \cite{nwaqar2022computation}. By default, we deploy four L-UAVs and one H-UAV server with maximum velocity of 25 m/s \cite{zyang2022dynamic}. To ensure uniform spatial distribution, the initial positions of four L-UAVs are set at (250, 250), (750, 250), (750, 750), and (250, 750), while the H-UAV is initially positioned at the center of entire area. Following \cite{myan2024edge, sli2024joint, hyuan2024cost}, the setup of simulation parameters are summarized in Table~\ref{tab:table2}.

\vspace{-0.4cm}
\begin{table}[!h]
\caption{Simulation Parameters}
\label{tab:table2}
\centering
\begin{tabular}{|m{1.5cm}<{\centering}|m{2.0cm}<{\centering}|m{1.5cm}<{\centering}|m{2.0cm}<{\centering}|}
\hline
\textbf{Parameter} & \textbf{Value} & \textbf{Parameter} & \textbf{Value}\\
\hline
$V$ & [10, 40] & $B_{v,u}$ & 2 MHz\\
\hline
$D_v$ & [1, 10] Mb & $B_{u,H}$ & 10 MHz\\
\hline
$C_v$ & [10, 100] & $E_q$ & 4 J\\
\hline
$\tau_{v,max}$ & [50, 200] ms & $e_m$ & 0.5 J\\
\hline
$F_u$ & 10 GHz & $k_u$ & $10^{-27}$\\
\hline
$F_H$ & 50 GHz & $M_u$ & 4 Kg\\
\hline
$P_v$ & 0.5 W & $d_{safe}$ & 5 m\\
\hline
$P_u$ & 1 W & $H_1$ & 100 m\\
\hline
$\gamma_0$ & -50 dB & $H_2$ & 150 m\\
\hline
$N_0$ & -174 dBm/Hz & $\tau$ & 0.2 s\\
\hline

\end{tabular}
\end{table}

\subsection{Performance Comparison with Benchmarks}

We conduct performance comparison of the proposed LATUS method with the following benchmark methods:

\begin{itemize}
    \item \textbf{LATUS with Fixed H-UAV Trajectory (FT-LATUS)}: The H-UAV in proposed method is modified to follow a fixed trajectory traversing along the diagonal of the field.
    \item \textbf{HAP-UAV Collaborative Resource Allocation Algorithm (HURA)}: The HAP-UAV collaborative resource optimization algorithm in \cite{sli2024joint} minimizing task delay with fixed energy constraint in each time slot.
    \item \textbf{UAV-Terrestrial Server Collaborative Delay-Centric Algorithm (UTDC)}: The UAV-terrestrial server collaborative task delay minimization method in \cite{yliu2024latency} without energy limitation.
    \item \textbf{HAP-UAV Collaborative Energy-Centric Algorithm (HUEC)}: UAV energy consumption minimization algorithm in \cite{hli2023energy} under task delay constraints with HAP employed as backup server.
\end{itemize}

\begin{figure}[!t]
\centering
\includegraphics[width=2.9 in]{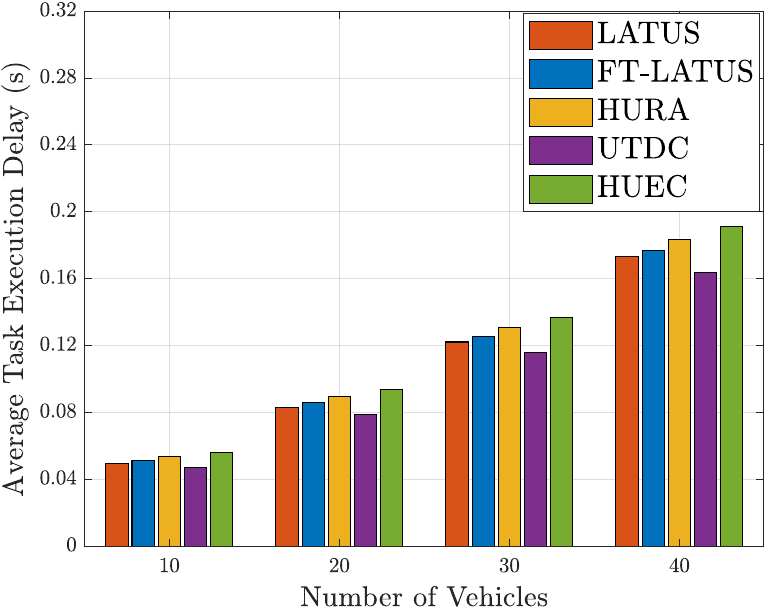}
\caption{The comparison of average task execution delay among different methods with varying numbers of vehicles.}
\label{bsl_delay}
\end{figure}

\emph{(1) Average task execution delay with varying numbers of vehicles:} Fig.~\ref{bsl_delay} illustrates the average task execution delay for different methods with varying numbers of vehicles. Given that merely minimizing task delay is not the dominant strength of LATUS, Fig.~\ref{bsl_delay} indicates that it can achieve comparable or even lower time-average task delay than benchmarks on the basis of flexibly trading off between task delay and L-UAV energy consumption. LATUS reduces task delay by around 12\% compared with HUEC, since HUEC is devoted to minimizing L-UAV energy consumption while merely satisfying task delay requirement. Compared to HURA with fixed energy limit in each time slot, LATUS will fully commit to minimizing task delay when there is no L-UAV energy deviation, thereby achieving approximately 9\% reduction in time-average task delay. Since UTDC solely prioritizes task delay minimization, its average task delay is inevitably lower than that of LATUS, yet it is only reduced by up to 5\%. The unoptimized H-UAV trajectory in FT-LATUS increases the LU2HU transmission delay and energy, resulting in roughly 3\% increment of total task delay compared with LATUS.

\begin{figure}[!t]
\centering
\includegraphics[width=2.9 in]{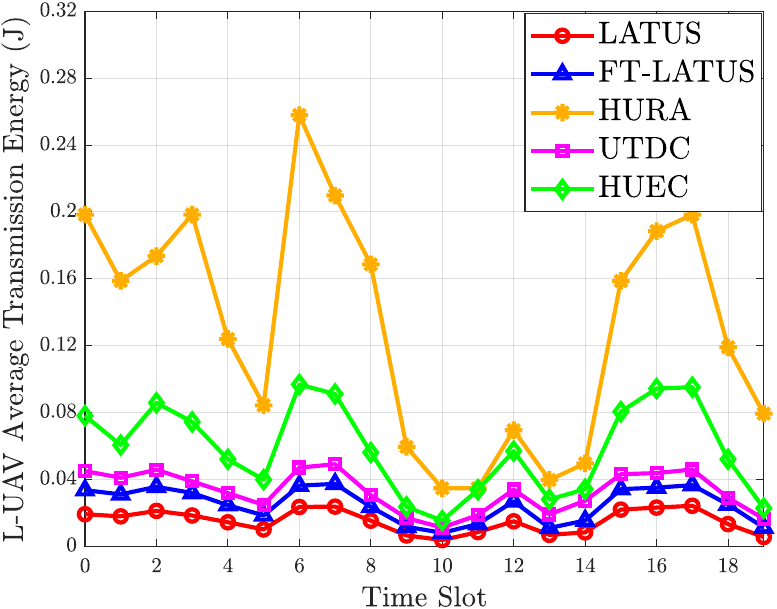}
\caption{The comparison of average L-UAV transmission energy among different methods.}
\label{tx_energy}
\end{figure}

\emph{(2) Average L-UAV transmission energy:} Fig.~\ref{tx_energy} shows the average L-UAV transmission energy among different methods over time slots. LATUS shortens the communication distance between L-UAVs and H-UAV, proactively adjusts data size transmitted by L-UAVs guided by their real-time energy states, and jointly optimizes trajectories of two-tier UAVs to enhance their connectivities. These coordinated mechanisms collectively contribute to its minimum L-UAV transmission energy. Compared with FT-LATUS, where unrefined trajectory of H-UAV prevents it from balancing communications with multiple L-UAVs in various conditions, LATUS attains 26\% reduction on average. UTDC employs ground base station as backup server. The complicated ground environment and higher probability of non-LoS links yield degraded channel quality. Moreover, the fixed location of ground server not only enlarges the communication distance to L-UAVs compared with H-UAV, but also precludes the dynamic repositioning that allows backup server to approach energy-critical or task-heavy L-UAVs. These shortcomings raise the average transmit energy of L-UAVs by 44\% relative to LATUS. As for HURA and HUEC, the excessive operating altitude of HAP significantly lengthens the links to low-altitude UAVs, subjecting the communications to severe path loss. Consequently, the average transmit energy of L-UAVs in HURA and HUEC are approximately 7 and 4 times higher than that of LATUS, respectively. HUEC places greater emphasis on L-UAV energy conservation than HURA, but its L-UAV transmit energy remains markedly higher than that achieved by the proposed scheme. This further underscores the advantage of low-altitude tiered UAV architecture.

\begin{figure}[!t]
\centering
\includegraphics[width=2.9 in]{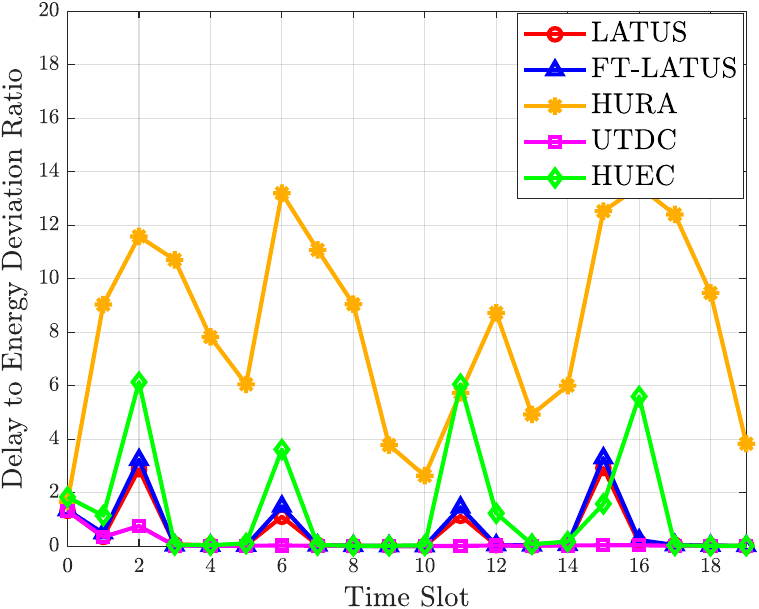}
\caption{The comparison of the delay to energy deviation ratio among different methods.}
\label{dedr}
\end{figure}

\emph{(3) Delay to energy deviation ratio:} To reflect the flexible tradeoff between task delay and L-UAV energy consumption, we define the Delay to Energy Deviation Ratio (DEDR) as the ratio of average task delay to average energy quota deviation of L-UAVs (plus a tiny constant to avoid zero-division). An increase in energy deviation indicates a higher degree of energy shortage, hence LATUS will trade task delay for conserving L-UAV energy, and vice versa. Therefore, a stabler DEDR implies a better tradeoff control. Fig.~\ref{dedr} illustrates that LATUS exhibits the most stable DEDR across time slots. FT-LATUS undergoes a slightly larger fluctuation, with an average increase of 15\%, since the unrefined H-UAV trajectory results in marginally reduced energy stability. The fluctuation of HURA is much more pronounced, 14 times higher than LATUS on average, because task delay fluctuates sharply with varying task volume due to the strict single-slot energy constraint of L-UAVs. Likewise, the L-UAV energy deviation of HUEC increases with a surge in task volume and vice versa, leading to a fluctuation roughly 4 times larger than LATUS. UTDC solely minimizes task delay without energy limits, leading to a continuous accumulation of L-UAV energy deviation. The progressively increasing denominator causes its DEDR to gradually approach zero.

\begin{figure}[!t]
\centering
\includegraphics[width=3.0 in]{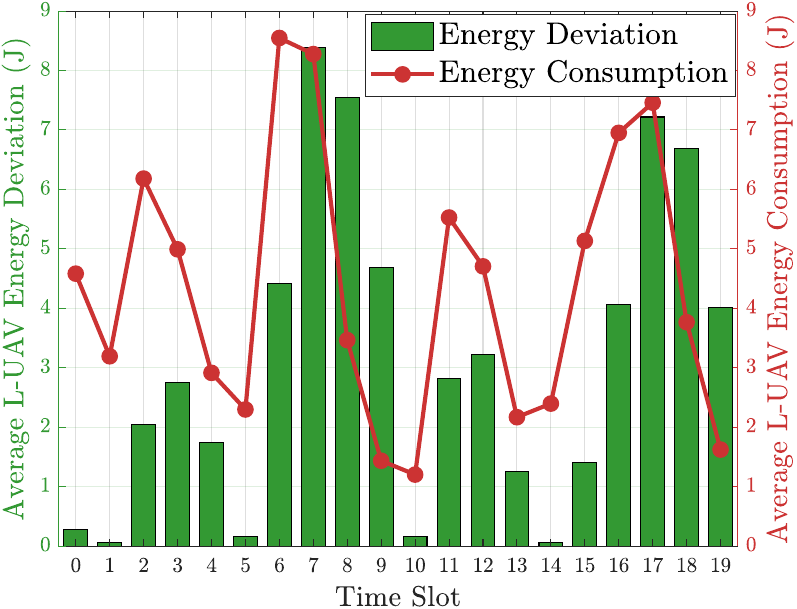}
\caption{Impact of energy quota deviation on L-UAV energy consumption.}
\label{deviation}
\end{figure}

\vspace{-0.1cm}
\subsection{Variation of L-UAV Energy Consumption with Energy Quota Deviation}

Fig.~\ref{deviation} illustrates the trend of average L-UAV energy consumption with their average energy quota deviation. Note that each green bar represents the average L-UAV energy deviation at the end of each time slot, which also serves as the one at the beginning of the next time slot. On one hand, a large energy quota deviation of L-UAVs indicates a higher level of energy shortage. LATUS will drive the system to reduce their energy expenditure. Consequently, it can be observed that severe energy deviations can lead to reduced average L-UAV energy consumption in subsequent time slots (e.g., time slots 7, 11, and 17). On the other hand, Fig.~\ref{deviation} also confirms that LATUS can maintain L-UAV energy quota deviation in a dynamically stable state.

\begin{figure}[!t]
\centering
\includegraphics[width=3.4 in]{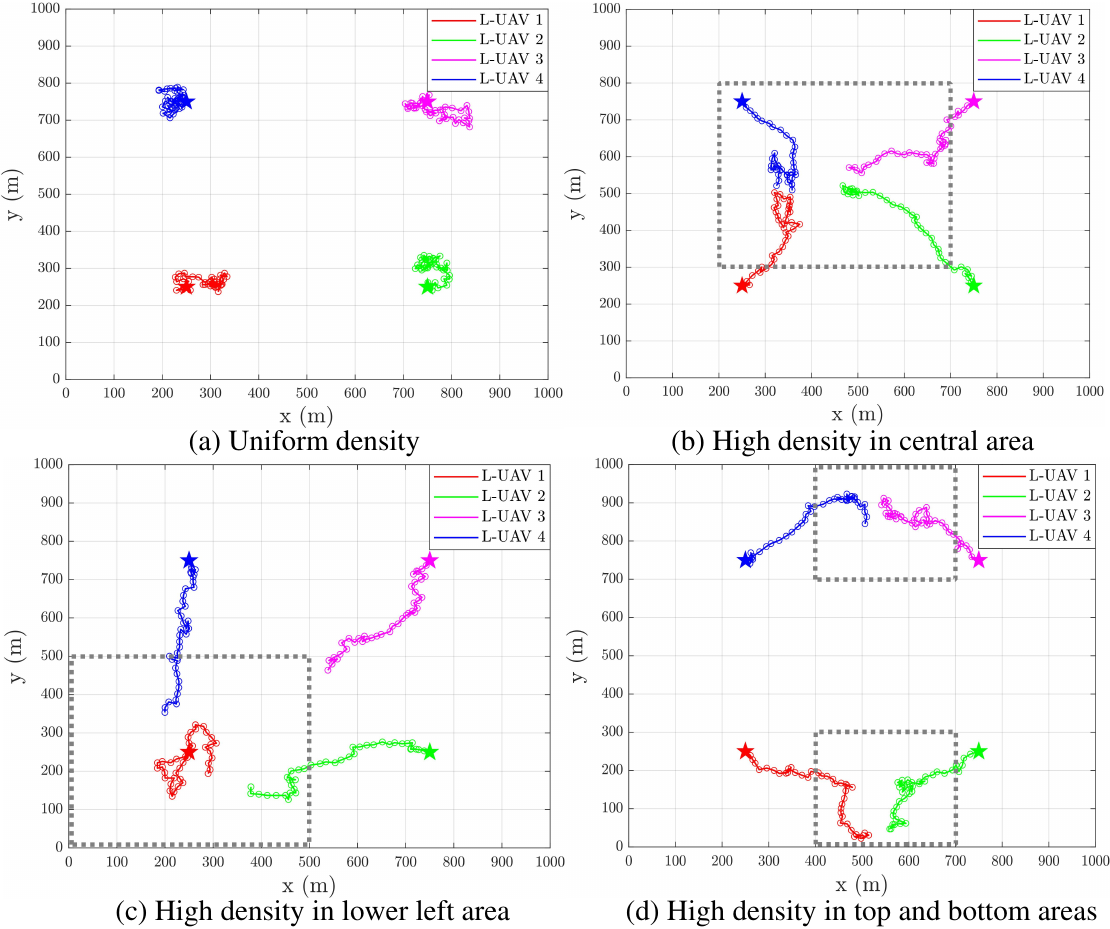}
\caption{Trajectories of L-UAVs under various vehicle distributions.}
\label{trajectory}
\end{figure}

\subsection{Trajectories of L-UAVs under Varying Vehicle Distributions}

Fig.~\ref{trajectory} shows the trajectories of L-UAVs under various distributions of requesting vehicles. The initial positions and trajectories of the four L-UAVs deployed by default are indicated by pentagrams and curves in distinct colors, respectively. In Fig.~\ref{trajectory} (a), vehicles are uniformly distributed in the test area, so the four L-UAVs wander and work near their respective initial positions to balance the communications with multiple served vehicles. In Fig.~\ref{trajectory} (b), (c), and (d), vehicle density is concentrated in the central, lower-left, and top and bottom areas, respectively, indicated by the gray dashed-line boxes. The primary goal of trajectory optimization for L-UAVs is to reduce V2LU communication delays by shortening their distances to served vehicles. Hence, it is evident that L-UAVs can adaptively fly towards vehicle-dense regions to provide computing services while avoiding collisions. Furthermore, as depicted in Fig.~\ref{trajectory} (c), L-UAV 1 is positioned closest to vehicle-dense area and shoulders the heaviest workload. Consequently, LATUS prioritizes reducing its flight distance to conserve more flight energy for computation and communication. Additionally, upon arriving at vehicle-dense zones, L-UAVs spontaneously reconfigure their spatial distribution toward dynamic uniformity, thereby facilitating inter-UAV load balancing. These results confirm the dynamic adaptability of LATUS to spatiotemporal variations in computational demands and its systemic stability.

\begin{figure}[!t]
\centering
\includegraphics[width=2.9 in]{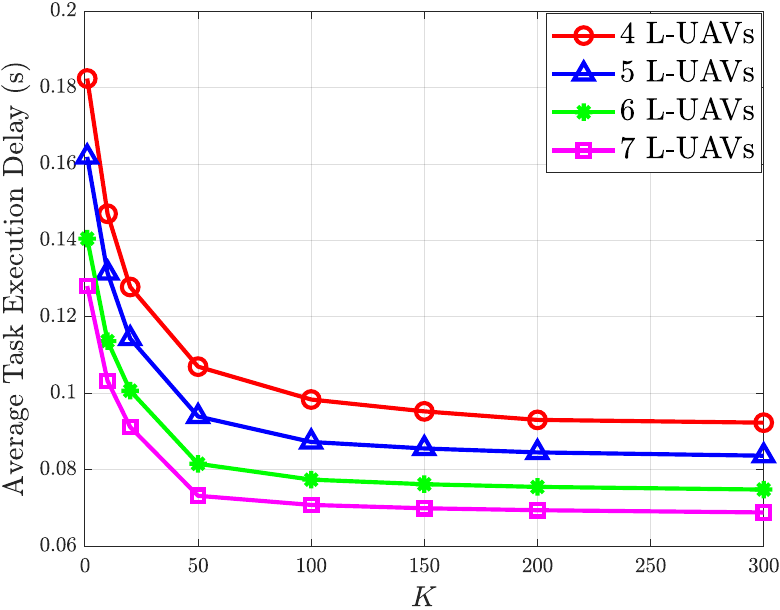}
\caption{Impact of number of L-UAVs and Lyapunov control factor $K$.}
\label{num_k}
\end{figure}

\subsection{Impact of Number of L-UAVs and Factor $K$}

Fig.~\ref{num_k} illustrates impact of the number of L-UAVs and the Lyapunov factor $K$ on average task execution delay. An increased number of L-UAVs implies a richer computing resources in this UAV system, resulting in a lower average task delay. For a given number of L-UAVs, as the factor $K$ increases, LATUS places higher priority on minimizing delay, which progressively reduces the average task execution delay. When $K$ keeps increasing and reaches a certain level, the average task delay asymptotically approaches that of the method focusing solely on delay minimization.

\begin{figure}[!t]
\centering
\includegraphics[width=2.8 in]{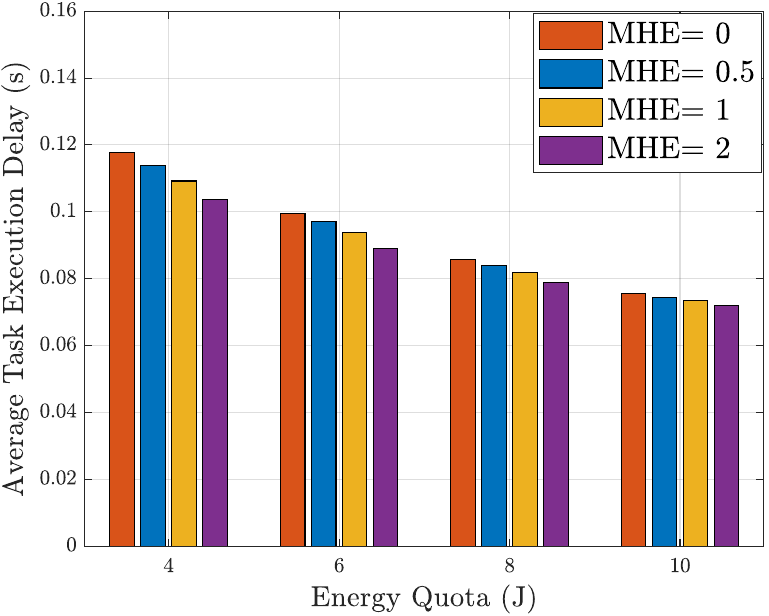}
\caption{Impact of energy quota and maximum harvestable energy.}
\label{mhe}
\end{figure}

\subsection{Impact of Energy Quota and Maximum Harvestable Energy}

Fig.~\ref{mhe} shows the impact of the L-UAV energy quota and the maximum harvestable energy (MHE) per time slot on average task execution delay. Overall, a larger energy quota and MHE enable L-UAVs to have more sufficient energy supply for providing computing services in each time slot, which is conducive to reducing average task delay. Specifically, when the L-UAV energy quota is relatively scarce, the amount of MHE becomes particularly valuable and has a more pronounced effect on average task delay. As the energy quota becomes increasingly ample, the impact of MHE becomes less significant. This highlights that applying energy harvesting technology can effectively mitigate the energy limitations of lightweight UAVs and accelerate task processing, especially when the energy quota is tight. A further insight is that energy harvesting technology also holds the potential to reduce the required number of UAVs, thereby saving the deployment cost of the whole system.


\section{Conclusions}

This paper proposed a multi-tier UAV edge computing system to minimize task delay while ensuring the long-term energy stability of L-UAVs. By leveraging Lyapunov optimization, we decoupled the long-term problem into manageable online sub-problems, dynamically balancing task delay and L-UAV energy consumption. Our solution jointly optimized vehicle to L-UAV matching, task assignment, resource allocation, and trajectory control. Simulations confirmed that our approach outperforms benchmarks, achieving superior delay-energy stability and comparable latency, providing a foundational framework for stable and efficient hierarchical UAV edge computing.

 




\vfill

\end{document}